\documentclass{aastex631}
\usepackage{graphicx}

\newcommand\nasagoddard{
NASA Goddard Space Flight Center, Greenbelt, Maryland, USA
}

\newcommand{\PSUAA}{Department of Astronomy \& Astrophysics, 525 Davey Laboratory, Penn State, University Park, PA, 16802, USA}
\newcommand{\PSUCEHW}{Center for Exoplanets and Habitable Worlds, 525 Davey Laboratory, Penn State, University Park, PA, 16802, USA}

\newcommand{\PSETI}{Penn State Extraterrestrial Intelligence Center, 525 Davey Laboratory, Penn State, University Park, PA, 16802, USA}

\begin{document}

\title{HZ\_evolution: A Package to Calculate Habitable Histories}
\author[0000-0003-3989-5545]{Noah Tuchow}
\affil{\nasagoddard}

\author[0000-0001-6160-5888]{Jason Wright}
\affil{\PSUAA}
\affil{\PSUCEHW}
\affil{\PSETI}

\begin{abstract}

We present \texttt{HZ\_evolution}, a Python package to characterize the habitable histories of exoplanets. Given inputs of a planet's current effective flux and host star properties, \texttt{HZ\_evolution} calculates its instellation history, the evolution of the star's Habitable Zone, and the duration the planet spends inside or outside the Habitable Zone.
\end{abstract}

\section{Habitable Histories}

In the search for biosignatures, it makes sense to consider the amount of time a planet spends in the Habitable Zone (HZ). The current climate and habitability of a planet depend on its past evolution, so one should consider whether a planet has hosted liquid water for a long enough duration for the development of life and the emergence of biosignatures. 

Modeling the climate evolution of exoplanets is complex and multifaceted, depending on many factors such as the planet's geophysical evolution, atmospheric evolution, and instellation history. Of these influences, we can expect that only the evolution of the planet's host star will be well understood for newly detected exoplanets. With knowledge of how a star evolves in luminosity and effective temperature, we can understand how the position of its HZ and the amount of stellar flux the planet receives, $S_{\mathrm{eff}}$, change in time. Despite limited knowledge about a planet, the instellation history provides invaluable information for one of the key drivers of climate evolution. 

Our \texttt{HZ\_evolution} python package\footnote{\url{https://zenodo.org/records/13840792}} calculates a planet's instellation history and the evolution of its star's HZ \citep{HZevolution}.  Building off the \texttt{CHZ\_calculator} of \citet{Tuchow2023}, \texttt{HZ\_evolution} takes a user-specified stellar evolutionary model and HZ formulation and calculates the extent of the Continuously Habitable Zone (CHZ); the time a planet has spent within the HZ (i.e.\ the \textit{habitable duration}, $\tau$); and the time spent interior to and exterior to the HZ ($t_\mathrm{int}$ and $t_\mathrm{ext}$ respectively) \citep{Tuchow2020,Tuchow2023}. 
\texttt{HZ\_evolution} provides integration with the \texttt{isochrones} package, allowing users to generate evolutionary tracks from the MIST stellar model grid \citep{Morton2015,Dotter2016,Choi2016}. 
To allow the functions to be evaluated rapidly, 
\texttt{HZ\_evolution} can generate grids of $\tau$, $t_\mathrm{int}$, and $t_\mathrm{ext}$ as a function of stellar fundamental properties (Mass, age, and metallicity) and $S_\mathrm{eff}$, which can be interpolated over. This allows \texttt{HZ\_evolution} to be integrated within MCMC methods to propagate model uncertainties to these calculated quantities.

\section{Example Case: Kepler 62\textit{f}}

\begin{figure}
    \plotone{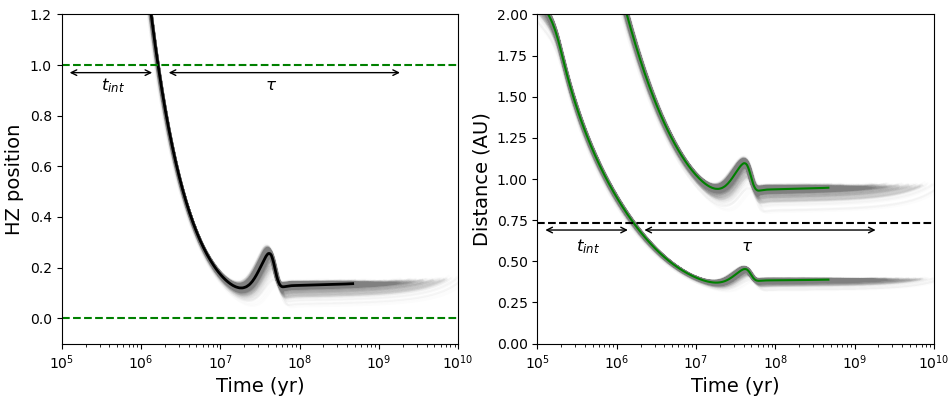}
    \caption{Habitable history of Kepler 62\textit{f}, incorporating uncertainties in host star properties. \textit{Left:} Instellation history of the planet in terms of normalized HZ flux. Dashed green lines represent the HZ. \textit{Right:} Evolution of the HZ boundaries. The black dashed line represents the position of the planet.} 
    \label{instellation_hist}
\end{figure}

 As an example application we consider Kepler 62\textit{f}, a roughly 1.4 $R_\oplus$ planet orbiting in the HZ of a late G or early K dwarf star with a period of 267.29 days \citep{Borucki2013}.
Using observed properties of Kepler 62 from \citet{Borucki2018} and \citet{Fulton2018} we fit a MIST model to the star. 
We obtain a best fit solution for Kepler 62 with the following stellar properties: [Fe/H]=$-0.35^{+0.07}_{-0.07}$, $M = 0.73^{+0.01}_{-0.01} M_\odot$, and log(age /\mbox{yr}) = $8.59^{+0.70}_{-0.59}$. Because it is a relatively low mass star, isochrones can only roughly constrain its age, while its mass and metallicity are better known. 
Note that our goal is not to precisely obtain the fundamental stellar properties of Kepler 62, but rather to illustrate how \texttt{HZ\_evolution} can be incorporated into MCMC fitting to calculate habitable histories with realistic uncertainties.

For each point in the MCMC chain, we interpolate over a precomputed grid of habitable durations calculated as a function of $S_\mathrm{eff}$, stellar mass, EEP (equivalent evolutionary phase, a proxy for age), and [Fe/H]. This grid is calculated using MIST evolutionary tracks and the \citet{Kopparapu2013} optimistic HZ. We find a habitable duration of $\log(\tau/ \mathrm{yr})=8.61^{+0.68}_{-0.59}$, which is about as well constrained as the age of the star. This makes sense because $\tau$ is closely correlated with stellar age. Interestingly the time interior to the HZ is $t_\mathrm{int}=1.67^{+0.01}_{-0.02}$ Myr, which is much better constrained than $\tau$ because $t_\mathrm{int}$ is most sensitive to a star's pre-main sequence evolution, which depends more strongly on the stellar mass than the current age. 
 $t_\mathrm{ext}$ is unambiguously zero since no points in the MCMC chain have the planet exterior to the HZ.

Figure \ref{instellation_hist} shows how the planet's position in the HZ changes as a consequence of the changing stellar flux, or equivalently how the spatial position of the HZ evolves. In the left panel we plot the normalized HZ flux $x= (S_\mathrm{eff}- S_\mathrm{outer}) / (S_\mathrm{inner}-S_\mathrm{outer})$, where $S_\mathrm{inner}$ and $S_\mathrm{outer}$ are the inner and outer HZ flux boundaries, corresponding to values of $x=1$ and $x=0$  respectively.  We plot this quantity rather than flux because the HZ position differs at each point in the MCMC chain; using the normalized HZ flux allows the instellation histories to be directly comparable across the samples. The right panel represents the same information expressed in terms of the spatial position of the HZ. This plot of the instellation history closely tracks the luminosity evolution of the host star, with luminosity falling rapidly on the Hayashi track, slightly increasing on the Henyey track, and finally settling on the main sequence where luminosity gradually increases. 
Here instellation is calculated at a fixed position around the star, starting at the beginning of the star's evolutionary track. However, during the star's early history the planet may still be forming, and may migrate to its current position. To account for this, \texttt{HZ\_evolution} includes an option to calculate habitable durations for a later starting time, $t_0$, such as a timescale for planet formation or volatile delivery. In our example, we don't make any assumptions about $t_0$, but for a more realistic estimate of $t_0=50$ Myr, $t_{int}$ goes to zero and the planet would sustain habitable conditions from the starting time to the current age.

In both panels, the best fit model is plotted as a solid line and 1000 random draws from MCMC chains are plotted in gray. This illustrates the uncertainty in the instellation history and allows one to envision the habitable history as either a fuzzy curve in instellation as a function of time, or alternatively as fuzzy boundaries of the HZ evolving in time for a planet at fixed separation. 
The habitable duration and time spent interior to the HZ can be easily inferred from either panel of Figure \ref{instellation_hist} by looking at the intercept of the planet's position with the HZ boundaries, as labeled in each panel.  This example indicates that, even if the current age cannot be well constrained, one can gain an accurate picture of a planet's instellation history using the \texttt{HZ\_evolution} package. We can see that the past history of the planet, such as when it first entered the HZ, is well constrained and doesn't vary substantially between different points in the MCMC chain. 

This example demonstrates that the \texttt{HZ\_evolution} package is a powerful tool that can assess the instellation histories of planets in the HZ. \texttt{HZ\_evolution} is publicly available at \url{https://zenodo.org/records/13840792}. 

\section*{Acknowledgements}
NWT is supported by an appointment with the NASA Postdoctoral Program at the NASA Goddard Space Flight Center, administered by Oak Ridge Associated Universities under contract with NASA.  The Center for Exoplanets and Habitable Worlds 
and the Penn State Extraterrestrial Intelligence Center are
supported by Penn State and its Eberly College of Science.

\bibliographystyle{aasjournal}
\bibliography{sources}

\end{document}